\documentclass[aps]{revtex4}

\usepackage[small]{caption}
\usepackage{amsmath,amsfonts,amscd,amssymb,epsf, epsfig,mathrsfs,graphicx,color}

\newcommand{\pa}{\partial}

\begin{document}

\title{Electromagnetic Casimir effect on the boundary of a $\boldsymbol{D}$-dimensional cavity and the high temperature asymptotics}
\author{L.P. Teo}\email{ LeePeng.Teo@nottingham.edu.my}\address{Department of Applied Mathematics, Faculty of Engineering, University of Nottingham Malaysia Campus, Jalan Broga, 43500, Semenyih, Selangor Darul Ehsan, Malaysia. }

\begin{abstract}
We consider the finite temperature Casimir stress acting on the boundary of a $D\geq 3$ dimensional cavity due to the vacuum fluctuations of electromagnetic fields. Both perfectly conducting and infinitely permeable boundary conditions are considered, and it is proved that they correspond mathematically to the relative and absolute boundary conditions. The divergence terms of the Casimir free energy are related to the heat kernel coefficients of the Laplace operator. It is shown that the Casimir stress is free of divergence if and only if $D$ is exactly three. The high temperature asymptotics of the regularized Casimir free energy are also found to depend on the heat kernel coefficients. When $D>3$, renormalization is required to remove terms of order higher than or equal to $T^2$.
\end{abstract}
 \maketitle

 \section{Introduction}

Casimir effect has aroused the interest of a lots of theoretical physicists and mathematicians. In the seminal work \cite{12}, Casimir predicted an attractive force acting between two parallel perfectly conducting plates due to the vacuum fluctuations of electromagnetic fields. Advances in experiments have confirmed the existence of this effect (see e.g. \cite{14} and the references therein). However, Casimir self-energies remain    elusive. Contrary to the pervasive belief at that time that Casimir stress is always attractive, Boyer \cite{13} showed in contrary that the Casimir stress acting on a perfectly conducting spherical shell is repulsive. This work has been extended to the finite temperature case in \cite{15} and it is proved that the Casimir stress remains repulsive at finite temperature.

In general, the computation of Casimir self-energy is not a simple task. By definition, the zero temperature Casimir energy is defined as the sum of ground state energies, which is generically divergent. Several well-defined regularization schemes have been widely adopted such as exponential cut-off method and zeta regularization. However, for the Casimir stress acting on the boundary of a cavity, it is natural to expect some cancelations of divergences of the self-energies inside and outside the cavity, which may render the Casimir stress acting on the boundary of the cavity finite without any regularization. This has proved to be the case in \cite{7} for a cavity in a (3+1)-dimensional Minkowski spacetime. A natural question to ask is what happens in higher dimensional spacetime? One of the main purpose of this article is to address this question.

One of the questions arises when considering electromagnetic field in higher dimensional spacetime is the natural boundary conditions to be imposed. In \cite{2}, the perfectly conducting and infinitely permeable boundary conditions in (3+1)-dimensional spacetime have been extended to higher dimensional spacetime. When considering electromagnetic Casimir effect on a higher dimensional spherical shell, we \cite{10} have observed that for spheres, the perfectly conducting and infinitely permeable boundary conditions are equivalent respectively to the relative and absolute boundary conditions for one-forms defined in mathematics literature \cite{3,1}. In this work, we show that these equivalences hold for any geometric configurations. As a result, we can apply well-known results about heat kernel coefficients for differential forms with relative or absolute boundary conditions  to study the divergence structure of Casimir effect on the boundary of a $D$-dimensional cavity. Finally, we discuss the high temperature asymptotic expansion of the Casimir energy and relate it to the heat kernel coefficients.

\section{Electromagnetic field in a $\boldsymbol{(D+1)}$-dimensional spacetime and the boundary conditions}\label{sec2}
Consider a bounded region $M$ with boundary $B=\pa  M$ in a $D$-dimensional space, not necessary the standard Euclidean space. Assume that $M$ is connected and the boundary $B$ is  smooth. Let $\text{x}=(x^{a})$  be a coordinate system on $M$. Assume that  the metric on the spacetime $\mathbb{R}\times M$ has the form
$$ds^2=g_{\mu\nu}dx^{\mu}dx^{\nu}=dt^2-\bar{g}_{ab}dx^adx^b.$$
Here $\mu$ and $\nu$ are indices running from $0$ to $D$, and $a$ and $b$ are indices running from $1$ to $D$.  The strength of an electromagnetic field is represented by a two-form $F=F_{\mu\nu}dx^{\mu}dx^{\nu}$ which is the exterior derivative of a one-form $A=A_{\mu}d x^{\mu}$, namely, $F_{\mu\nu}=\pa_{\mu}A_{\nu}-\pa_{\nu}A_{\mu}$. The equation of motion for the electromagnetic field is
$$\delta F=\frac{1}{\sqrt{|g|}}\frac{\pa}{\pa x^{\nu}}\left(\sqrt{|g|}F^{\nu\mu}\right)=0,$$
where $\delta$ is the co-differential operator.
The one-form $A$ is defined up to an exact differential, i.e., $A$ and $A+d\phi$ define the same field $F$. To eliminate the gauge degree of freedom, we   impose the radiation gauge where $A_t=0$ and
\begin{equation}\label{eq9_28_1}
\delta A=\frac{1}{\sqrt{|g|}}\frac{\pa}{\pa x^{\mu}}\left(\sqrt{|g|}A^{\mu }\right)=0.
\end{equation}
As usual, assume that the field is monochromatic, i.e.,
$$F_{\mu\nu}(t,x^a)=\text{F}_{\mu\nu}(x^a)e^{-it\omega}.$$
Then we can also write $A$ as
$$A_{\mu}(t,x^a)=\text{A}_{\mu}(x^a)e^{-it\omega}.$$
With the condition $A_t=0$, we can regard $\text{A}_{a}dx^{a}$ as a one-form on $M$. The gauge condition $\delta A=0$ is equivalent to $\delta\text{A}=0$, i.e., A is a co-closed one-form on $M$. The equation of motion can be written as
$(d\delta+\delta d)A=0$,
which is equivalent to $$\Delta \text{A}=\omega^2\text{A},$$
i.e., A is an eigen-one-form of the Laplace operator on $M$ with eigenvalue $\omega^2$.

For electromagnetic field in $(3+1)$-dimensional Minkowski spacetime, there are two natural boundary conditions that one can impose on the boundary of an object: the perfectly conducting boundary condition  and the infinitely permeable boundary condition. These boundary conditions have been extended to general $(D+1)$-dimensional spacetimes as follows \cite{2}:  the perfectly conductor boundary condition is
\begin{equation}\label{eq4_5_1}\left.n^{\mu}(*F)_{ \mu\nu_1\ldots\nu_{D-2}}\right|_{ \text{boundary}}=0,\end{equation}and the infinitely permeable boundary condition is
\begin{equation}\label{eq4_5_2}\left.n^{\mu}F_{ \mu\nu }\right|_{\text{boundary}}=0.\end{equation}
Here $n^{\mu}$ is a unit normal vector to the boundary, and $*F$ is the dual tensor of $F$.

In studying spectra of differential forms on manifolds with boundaries, there are   two natural boundary conditions that have been considered: the absolute boundary conditions and the relative boundary conditions.   As in \cite{3,1}, treating $M$ as a manifold with boundary $B$, we can identify a neighborhood of the boundary with the collar $B\times [0, i(M))$, where $i(M)>0$ is the injectivity radius. Given $y\in  B$, let $r\mapsto (y,r)$ be the unit speed geodesic that is perpendicular to $B$ at $y$. Then $(y,r)$ defines a local coordinate system near the boundary of $M$, with metric
$$dr^2+\tilde{g}_{\alpha\beta}dy^{\alpha}dy^{\beta}.$$
Using this coordinate system, a one-form $\text{A}$ can be written as
$$\text{A}= \text{A}_rdr+ \text{A}_{\alpha}dy^{\alpha}$$
on  a neighborhood of $B$. For a one-form, the absolute boundary condition is defined as \cite{3}:
\begin{align}\label{eq10_17_1}\text{A}_r\bigr|_{\text{boundary}}=0\hspace{1cm}\text{and}\hspace{1cm}\pa_r\text{A}_{\alpha}\bigr|_{\text{boundary}}=0,
\end{align}
and the relative boundary condition is defined as \cite{3}:
\begin{align}\label{eq10_17_2}
\pa_r\left(\sqrt{|\tilde{g}|}\text{A}^r\right)\bigr|_{\text{boundary}}=0\hspace{1cm}\text{and}\hspace{1cm}\text{A}_{\alpha}\bigr|_{\text{boundary}}=0.
\end{align}

To the best of our knowledge, except for our work \cite{10}, no other work has ever explored the relations between the physically defined boundary conditions \eqref{eq4_5_1} and \eqref{eq4_5_2} and the mathematically defined boundary conditions \eqref{eq10_17_1} and \eqref{eq10_17_2}.
In the following, we want to show that the perfectly conducting boundary condition for $F$ is the same as the relative boundary condition for A, and the infinitely permeable boundary condition for $F$ is the same as the absolute boundary condition for A. These equivalences have been proved in our work \cite{10} for a $D$-dimensional ball.

First, consider the infinitely permeable boundary conditions. Using the local coordinates $x=(t,r,y)$, the metric is
 $$ds^2=g_{\mu\nu}dx^{\mu}dx^{\nu}=dt^2-dr^2-\tilde{g}_{\alpha\beta}dy^{\alpha}dy^{\beta}.$$ The infinitely permeable boundary condition \eqref{eq4_5_2} amounts to
$$F_{rt}\bigr|_{B}=0\hspace{1cm}\text{and}\hspace{1cm}F_{r\alpha}\bigr|_{B}=0.$$
Since
$$F_{rt}=-\pa_tA_r=i\omega \text{A}_re^{-i\omega t},$$
we find that $F_{rt}\bigr|_{B}=0$ if and only if $\text{A}_r\bigr|_{B}=0$.
On the other hand,
$$F_{r\alpha}=\left(\pa_r\text{A}_{\alpha}-\pa_{\alpha}\text{A}_r\right)e^{-i\omega t}.$$
Notice that $\text{A}_r\bigr|_{B}=0$ implies
$\pa_{\alpha}\text{A}_r\bigr|_{B}=0$. Hence, $F_{rt}\bigr|_{B}=0$ and $F_{r\alpha}\bigr|_{B}=0$ if and only if $\text{A}_r\bigr|_{B}=0$ and $\pa_r\text{A}_{\alpha}\bigr|_{B}=0$. This shows that the infinitely permeable boundary condition for $F$ is the same as the absolute boundary condition for A.

For the perfectly conducting boundary conditions \eqref{eq4_5_1}, it is equivalent to
$$F_{t\alpha}\bigr|_{B}=0\hspace{1cm}\text{and}\hspace{1cm}F_{\alpha\beta}\bigr|_{B}=0.$$
As above, we find that  $F_{t\alpha}\bigr|_{B}=0$ if and only if $\text{A}_{\alpha}\bigr|_{B}=0$. On the other hand, since
$$F_{\alpha\beta}=\left(\pa_{\alpha}\text{A}_{\beta}-\pa_{\beta}\text{A}_{\alpha}\right)e^{-i\omega t},$$ and $\text{A}_{\alpha}\bigr|_{B}=0$ implies that
$\pa_{\beta}\text{A}_{\alpha}\bigr|_{B}=0$, therefore $F_{t\alpha}\bigr|_{B}=0$ implies $F_{\alpha\beta}\bigr|_{B}=0$. In this case, it seems that we do not get the condition $\pa_r\left(\sqrt{|\tilde{g}|}\text{A}^r\right)\bigr|_{B}=0$ for relative boundary condition. However, notice that the gauge condition \eqref{eq9_28_1} implies that
\begin{equation*}
\pa_r\left(\sqrt{|\tilde{g}|}\text{A}^r\right)+\pa_{\alpha}\left(\sqrt{|\tilde{g}|}\text{A}^{\alpha}\right)=0.
\end{equation*}Here we have used the fact that $\sqrt{|g|}=\sqrt{|\tilde{g}|}$. Since $\text{A}_{\alpha}\bigr|_{B}=0$ implies that $\pa_{\alpha}\left(\sqrt{|\tilde{g}|}\text{A}^{\alpha}\right)\bigr|_{B}=0$, we see that the perfectly conducting boundary condition implies $\pa_r\left(\sqrt{|\tilde{g}|}\text{A}^r\right)\bigr|_{B}=0$. Thus perfectly boundary condition is equivalent to the relative boundary condition.

\section{Casimir free energy inside a $\boldsymbol{D}$-dimensional cavity}\label{sec3}
In this section, we give a review about the relations between the Casimir free energy, the heat kernel coefficients and the zeta functions. This is not new as it has appeared in a number of works on Casimir effect.

The Casimir free energy of the electromagnetic field in a bounded region $M$ with perfectly conducting or infinitely permeable boundary conditions is defined as
\begin{align}
\label{eq10_17_3}E_{\text{Cas}}(M;b)=\frac{1}{2}\sum_{\omega_{j;b}\neq 0}\omega_{j;b}+T\sum_{\omega_{j;b}\neq 0}\ln\left(1-e^{-\omega_{j;b}/T}\right).
\end{align}
 The first part is the zero temperature Casimir energy and the second part is the thermal correction.    $\omega_{j;b}^2$ are the  eigenvalues of the Laplace operator on one-forms A on $M$ which are co-closed, and subject to certain boundary conditions $b$. For electromagnetic field with perfectly conducting boundary conditions, $b=r$, the relative boundary conditions. For electromagnetic field with infinitely permeable boundary conditions, $b=a$, the absolute boundary conditions. The zero temperature Casimir energy
\begin{align}
\label{eq10_17_4}E_{\text{Cas}}^{T=0}(M;b)=\frac{1}{2}\sum_{\omega_{j;b}\neq 0}\omega_{j;b}
\end{align}
 is generically divergent, and a conventional way to regularize this sum is to introduce an exponential cut-off:
\begin{align*}
E_{\text{Cas}}^{T=0}(M;b)=\frac{1}{2}\sum_{\omega_{j;b}\neq 0}\omega_{j;b} e^{-\lambda \omega_{j;b}},
\end{align*}and consider the limit $\lambda\rightarrow 0^+$.

Define the zeta function $\zeta_b(s)$ and the heat kernel $K_b(t)$ to be
\begin{equation*}
\zeta_b(s)=\sum_{\omega_{j;b}\neq 0} \omega_{j;b}^{-2s},\hspace{1cm}K_b(t)=\sum_{\omega_{j;b}\neq 0}e^{-t\omega_{j;b}^2}.
\end{equation*}
As $t\rightarrow 0^+$, one has (see e.g. \cite{1,11,5}):
\begin{equation}\label{eq10_17_10}
K_b(t)\sim \sum_{n=0}^{\infty} c_{n;b}t^{\frac{n-D}{2}},
\end{equation}where the coefficients $c_{n;b}$ are given by
\begin{equation*}
c_{n;b}=\text{Res}_{s=\frac{D-n}{2}}\left(\Gamma(s)\zeta_b(s)\right).
\end{equation*}
In particular,
\begin{equation*}
\begin{split}
c_{D;b}=&\zeta_b(0),\\
c_{D+1;b}=&-2\sqrt{\pi} \text{Res}_{s=-\frac{1}{2}} \zeta_b(s).
\end{split}\end{equation*}
The   coefficients $c_{n;b}$, with $0\leq n \leq D+1$,   play   important roles in the divergence behavior of the Casimir energy at zero temperature and the asymptotic behavior of the Casimir free energy at high temperature (see e.g. \cite{9,11,5}).
Using the inverse Mellin transform formula
\begin{equation}\label{eq10_17_6}
e^{-\alpha}=\frac{1}{2\pi i}\int_{c-i\infty}^{c+i\infty}dz\,\Gamma(z)\alpha^{-z},
\end{equation} we have
\begin{equation*}
E_{\text{Cas}}^{T=0}(M;b)=\frac{1}{2}\frac{1}{2\pi i}\int_{c-i\infty}^{c+i\infty} dz\,\Gamma(z)\lambda^{-z}\zeta_b\left(\frac{z-1}{2}\right).
\end{equation*}Taking the residues at $z=D+1-n$, $n=0,\ldots, D+1$, gives
\begin{equation*}
E_{\text{Cas}}^{T=0}(M;b)=\sum_{n=0}^{D-1}\frac{\Gamma(D+1-n)}{\Gamma\left(\frac{D-n}{2}\right)}c_{n;b}\lambda^{n-D-1}-\frac{\psi(1)-\ln\lambda}{2\sqrt{\pi}}c_{D+1;b}
+\frac{1}{2}\text{FP}_{s=-\frac{1}{2}}\zeta_b(s)+o(\lambda).
\end{equation*}From here we see that the coefficients $c_{n;b}$, $n=0,\ldots, D-1$ and $D+1$  are associated with the divergence when $\lambda\rightarrow 0^+$.
Using zeta regularization, the regularized zero temperature Casimir energy is defined as (see e.g. \cite{1,11,5}):
\begin{equation}\label{eq10_17_7}
\begin{split}
E_{\text{Cas}}^{\text{reg},T=0}(M;b)=&\frac{1}{2}\text{FP}_{s=-\frac{1}{2}}\zeta_b(s)-\frac{c_{D+1;b}}{4\sqrt{\pi}}\ln\mu^2\\
=&\frac{1}{2}\left(\text{FP}_{s=-\frac{1}{2}}\zeta_b(s)+[\ln\mu^2]\text{Res}_{s=-\frac{1}{2}}\zeta_b(s)\right),
\end{split}\end{equation}where $\mu$ is a normalization constant. This gives an unambiguous regularized Casimir free energy if and only if $c_{D+1;b}= 0$.

 For the Casimir free energy, consider the thermal zeta function
\begin{equation*}
\zeta_{T;b}(s)=\sum_{\omega_{j;b}\neq 0}\sum_{l=-\infty}^{\infty}\left(\omega_{j;b}^2+[2\pi lT]^2\right)^{-s}.
\end{equation*}Using the formula
$$\sum_{l=-\infty}^{\infty}\exp\left(-t[2\pi l T]^2\right)=\frac{1}{2\sqrt{\pi t} T}\sum_{l=-\infty}^{\infty}\exp\left(-\frac{1}{t}\frac{l^2}{4T^2}\right),$$
we have
\begin{equation}\label{eq10_17_9}
\begin{split}
\zeta_{T;b}(s)=&\frac{1}{\Gamma(s)}\sum_{\omega_{j;b}\neq 0}\sum_{l=-\infty}^{\infty}\int_0^{\infty}dt\, t^{s-1}\exp\left(-t\omega_{j;b}^2-t[2\pi l T]^2\right)\\
=&\frac{1}{\Gamma(s)}\frac{1}{2\sqrt{\pi} T}\sum_{\omega_{j;b}\neq 0}\sum_{l=-\infty}^{\infty}\int_0^{\infty}dt\, t^{s-\frac{1}{2}-1}\exp\left(-t\omega_{j;b}^2
-\frac{1}{t}\frac{l^2}{4T^2}\right)\\
=&\frac{\Gamma\left(s-\frac{1}{2}\right)}{\Gamma(s)}\frac{1}{2\sqrt{\pi}T}\zeta_b\left(s-\frac{1}{2}\right)+\frac{1}{\Gamma(s)}\frac{2}{\sqrt{\pi}T}\sum_{\omega_{j;b}\neq 0}\sum_{l=1}^{\infty}\left(\frac{l}{2T\omega_{j;b}}\right)^{s-\frac{1}{2}}K_{s-\frac{1}{2}}\left(\frac{l\omega_{j;b}}{T}\right).
\end{split}\end{equation}
It follows that
\begin{equation*}
\begin{split}
\zeta_{T;b}(0)=&-\frac{1}{T}\text{Res}_{s=-\frac{1}{2}}\zeta_b(s),\\
\zeta_{T;b}'(0)=&-\frac{1}{T}\left(\text{FP}_{s=-\frac{1}{2}}\zeta_b(s)+(2-2\ln 2)\text{Res}_{s=-\frac{1}{2}}\zeta_b(s)\right)-2\sum_{\omega_{j;b}\neq 0}\ln \left(1-e^{-\omega_{j;b}/T}\right).
\end{split}\end{equation*}
Hence, the Casimir free energy \eqref{eq10_17_3} is given by
\begin{equation*}
E_{\text{Cas}}(M;b)=\sum_{n=0}^{D-1}\frac{\Gamma(D+1-n)}{\Gamma\left(\frac{D-n}{2}\right)}c_{n;b}\lambda^{n-D-1}-\frac{\psi(1)-1+\ln2-\ln\lambda}{2\sqrt{\pi}}c_{D+1;b}-\frac{T}{2}
\zeta_{T;b}'(0)+o(\lambda).
\end{equation*}Using zeta regularization, the regularized Casimir free energy is defined as
\begin{equation}\label{eq10_17_8}
E_{\text{Cas}}^{\text{reg}}(M;b)=-\frac{T}{2}\left(\zeta_{T;b}'(0)+[\ln\tilde{\mu}]^2\zeta_{T;b}(0)\right),
\end{equation}where $\tilde{\mu}=2\mu/e$. We can rewrite the cut-off dependent Casimir free energy as
\begin{equation}\label{eq4_25_2}
E_{\text{Cas}}(M;b)=\sum_{n=0}^{D-1}\frac{\Gamma(D+1-n)}{\Gamma\left(\frac{D-n}{2}\right)}c_{n;b}\lambda^{n-D-1}-\frac{\psi(1) -\ln[\lambda\mu]}{2\sqrt{\pi}}c_{D+1;b}+E_{\text{Cas}}^{\text{reg}}(M;b)+o(\lambda).
\end{equation}

For the high temperature asymptotic behavior of the Casimir free energy, we only need to consider the temperature correction term
\begin{equation*}
\Delta_TE_{\text{Cas}}(M;b)=T\sum_{\omega_{j;b}\neq 0}\ln\left(1-e^{-\omega_{j;b}/T}\right)=-T\sum_{\omega_{j;b}\neq 0}\sum_{l=1}^{\infty}\frac{1}{l}\exp\left(-\frac{l\omega_{j;b}}{T}\right).
\end{equation*}
Using the formula \eqref{eq10_17_6}, we have
\begin{equation*}
\begin{split}
\Delta_TE_{\text{Cas}}=&-   \frac{1}{2\pi i}\int_{c-i\infty}^{c+i\infty}dz\,\Gamma(z) \zeta_R(z+1)T^{z+1}\zeta_b\left(\frac{z}{2}\right)\\
=&- \frac{1}{\sqrt{\pi}} \frac{1}{2\pi i}\int_{c-i\infty}^{c+i\infty}dz\,2^{z-1}T^{z+1}\Gamma\left(\frac{z+1}{2}\right) \zeta_R(z+1)\Gamma\left(\frac{z}{2}\right)\zeta_b\left(\frac{z}{2}\right).
\end{split}
\end{equation*}Evaluate the residues at $z=D-n$, $n=0,1,2,\ldots$, we find that as $T\rightarrow \infty$,
\begin{equation*}
\begin{split}
\Delta_T E_{\text{Cas}}(M;b)\sim &-\frac{1}{\sqrt{\pi}}\sum_{\substack{n\geq 0\\ n\neq D, D+1}}2^{D-n} \Gamma\left(\frac{D-n+1}{2}\right)\zeta_R(D-n+1) c_{n;b}T^{D-n+1}-T\left(\zeta_b(0)\ln T+\frac{1}{2}\zeta_b'(0)\right)\\
&-\Bigl(1+\psi(1)+\ln (2\pi T)\Bigr)\text{Res}_{s=-\frac{1}{2}}\zeta_b(s)-\frac{1}{2}\text{FP}_{s=-\frac{1}{2}}\zeta_b(s).
\end{split}
\end{equation*}Together with the zero temperature term and the  functional equation for Riemann zeta function
\begin{equation}\label{eq10_17_11}\Gamma\left(\frac{s}{2}\right)\zeta_R(s)=\pi^{s-\frac{1}{2}}\Gamma\left(\frac{1-s}{2}\right)\zeta_R(1-s),\end{equation} we have
\begin{equation}\label{eq10_17_12}
\begin{split}
  E_{\text{Cas}}^{\text{reg}}(M;b)\sim &-\frac{1}{\sqrt{\pi}}\sum_{n=0}^{D-1}2^{D-n} \Gamma\left(\frac{D-n+1}{2}\right)\zeta_R(D-n+1) c_{n;b}T^{D-n+1}-T\left(\zeta_b(0)\ln T+\frac{1}{2}\zeta_b'(0)\right)\\
&-\Bigl(1+\psi(1)+\ln (2\pi )+\ln( T/\mu)\Bigr)\text{Res}_{s=-\frac{1}{2}}\zeta_b(s)- \sum_{n=D+2}^{\infty}\frac{1}{(2\pi)^{n-D}} \Gamma\left(\frac{n-D}{2}\right)\zeta_R(n-D) c_{n;b}\frac{1}{T^{n-D-1}}.
\end{split}
\end{equation}
This asymptotic expansion can also be derived from \eqref{eq10_17_8}. As in \cite{6}, from the first line of \eqref{eq10_17_9} and \eqref{eq10_17_10}, we have
\begin{equation*}
\begin{split}
\zeta_{T;b}(s)\sim &\zeta_b(s)+\frac{2}{\Gamma(s)}\sum_{l=1}^{\infty}\int_0^{\infty}dt\,t^{s-1}\sum_{n=0}^{\infty}c_{n;b} t^{\frac{n-D}{2}} \exp\left(-t[2\pi l T]^2\right)\\
=&\zeta_b(s)+\frac{2}{\Gamma(s)}\sum_{n=0}^{\infty} c_{n;b}\frac{\Gamma\left(s+\frac{n-D}{2}\right)}{(2\pi T)^{2s+n-D}}\zeta_R(2s+n-D).
\end{split}
\end{equation*}
Hence,
\begin{equation*}
\begin{split}
\zeta_{T;b}(0)\sim &\zeta_b(0)-c_{D;b}+\frac{1}{2\sqrt{\pi}T}c_{D+1;b}=\frac{1}{2\sqrt{\pi}T}c_{D+1;b},\\
\zeta_{T;b}'(0)\sim &\zeta_b'(0)+2\sum_{\substack{n\geq 0\\ n\neq D, D+1}}c_{n;b}\frac{\Gamma\left( \frac{n-D}{2}\right)}{(2\pi T)^{ n-D}}\zeta_R( n-D)+2c_{D;b}\ln T
-\frac{c_{D+1;b}}{\sqrt{\pi }T}\Bigl(\psi(1)+\ln (4\pi T)\Bigr).
\end{split}
\end{equation*}
Substituting these into \eqref{eq10_17_8} give \eqref{eq10_17_12}.

From \eqref{eq10_17_12}, we see that the high temperature   leading term is of order $T^{D+1}$ and it depends on $c_{0;b}$. The subsequent terms  of order $T^{D},\ldots,T^2$   depend on $c_{1;b},\ldots,c_{D-1;b}$. These terms will be important when we consider renormalization of the Casimir free energy. The physical meaningful terms are the terms of order less than   $T^2$, which include the $T\ln T$ term.

\section{The heat kernel coefficients}\label{sec4}
In this section, we use the results of \cite{3,1,4}  to derive the expressions for the first three heat kernel coefficients $c_{0;b}, c_{1;b}$ and $c_{2;b}$.

Let $\{\mu_{j,p;b}\}$ be the eigenvalues of the Laplace operator on $p$-forms on $M$ with either absolute ($b=a$) or relative ($b=r$) boundary conditions.
As in \cite{3}, denote by $a_n(\Delta_{p;b})$ the heat kernel coefficients for Laplace operator on $p$-forms with boundary conditions $b$. More precisely, they are coefficients that appear in the asymptotic expansion of the heat kernel:
\begin{equation*}
\sum_{j}e^{-t\mu_{j,p;b}}\sim \sum_{n=0}^{\infty}a_n(\Delta_{p;b}) t^{\frac{n-D}{2}}\hspace{1cm}(t\rightarrow 0^+).
\end{equation*}
Each of these heat kernel coefficients can be expressed as a sum of an integral over the manifold and an integral over the boundary of the manifold.

The set of eigenvalues $\{\omega_{j;b}^2\}$ we consider in the previous section is a subset of $\{\mu_{j,1;b} \}$ consists of eigenvalues of the Laplace operator on co-closed one-forms. One can show that   the difference of between the set $\{\mu_{j,1;b}\}$ and the set $\{\omega_{j,b}^2\}$ is the set  $\{\mu_{j,0;b}\}$ of eigenvalues of Laplace operator on functions ($0$-forms).
It follows that
\begin{equation}\label{eq4_25_1}
\begin{split}
c_{n;b}=&a_n(\Delta_{1;b})-a_n(\Delta_{0;b}).
\end{split}
\end{equation}
As a side remark, for functions, absolute boundary condition is the same as Neumann boundary condition, and relative boundary condition is the same as Dirichlet boundary condition.

The formulas for $a_0(\Delta_{p;b}), a_1(\Delta_{p;b})$ and $a_2(\Delta_{p;b})$ have been obtained in \cite{3,1,4}. Let us first define some terms. As in \cite{3}, let $\{e_1,\ldots e_{D}\}$ be a local orthonormal frame for the tangent bundle $TM$. Near the boundary, we choose a frame so that $e_D$ is the inward pointing geodesic normal. Let $\nabla_{e_i} e_j=\Gamma_{ijk}e_k$ be the Christoffle symbols of the Levi-Civita connection on $M$. Define
\begin{equation*}
R_{ijkl}=\left\langle \left(\nabla_{e_i}\nabla_{e_j}-\nabla_{e_j}\nabla_{e_i}-\nabla_{[e_i,e_j]}\right)e_k,\;e_l\right\rangle.
\end{equation*}This is the curvature tensor.  On the standard sphere, $R_{1212}=-1$. The Ricci tensor is given by
\begin{equation*}
\rho_{ij}=\sum_{k=1}^DR_{ikkj},
\end{equation*}whereas the scalar curvature is
\begin{equation*}
\tau=\sum_{i=1}^D\rho_{ii}.
\end{equation*}If $M$ is a bounded sub-manifold of $\mathbb{R}^D$ with metric induced from the standard Euclidean metric, $\tau=0$.

Near the boundary,  the second fundamental form is defined as
\begin{equation*}
L_{ab}=\left\langle \nabla_{e_a} e_b,\;e_m\right\rangle=\Gamma_{abm}.
\end{equation*}

From \cite{3,1,4}, we have  the following results:
\begin{equation}\label{eq10_18_1}
\begin{split}
a_{0}(\Delta_{p;a})=&\frac{1}{(4\pi)^{\frac{D}{2}}} h(D,p)\text{vol}(M),\\
a_{1}(\Delta_{p;a})=&\frac{1}{4}\frac{1}{(4\pi)^{\frac{D-1}{2}}} d_0(D,p)\text{vol}(\pa M),\\
a_{2}(\Delta_{p;a})=&\frac{1}{6}\frac{1}{(4\pi)^{\frac{D}{2}}}h_0(D,p)\left\{\int_M\tau+2\int_{\pa M}\sum_{a=1}^{D-1}L_{aa}\right\},
\end{split}
\end{equation}
where
\begin{equation*}
\begin{split}
h(D,p)=&\frac{D!}{p!(D-p)!},\\
h_0(D,p)=&h(D,p)-6h(D-2,p-1),\\
d_0(D,p)=&h(D-1,p)-h(D-1,p-1).
\end{split}
\end{equation*}Here we understand that $h(D,p)=0$ for $p<0$ or $p>D$.

From \eqref{eq4_25_1} and \eqref{eq10_18_1}, we find that for absolute boundary conditions,
\begin{equation}\label{eq4_25_3}
\begin{split}
c_{0;a}=&\frac{(D-1)}{(4\pi)^{\frac{D}{2}}}\text{vol}(M),\\
c_{1;a}=& \frac{(D-3)}{4(4\pi)^{\frac{D-1}{2}}}  \text{vol}(\pa M),\\
c_{2;a}=&\frac{(D-7)}{6 (4\pi)^{\frac{D}{2}}} \left\{\int_M\tau+2\int_{\pa M}\sum_{a=1}^{D-1}L_{aa}\right\}.
\end{split}
\end{equation}
For relative boundary conditions, since (see e.g. \cite{3})
$$a_n(\Delta_{p;r})=a_n(\Delta_{D-p;a}),$$ we have
\begin{equation}\label{eq4_25_4}
\begin{split}
c_{0;r}=&\frac{(D-1)}{(4\pi)^{\frac{D}{2}}}\text{vol}(M),\\
c_{1;r}=&-\frac{(D-3)}{4(4\pi)^{\frac{D-1}{2}}}  \text{vol}(\pa M),\\
c_{2;r}=&\frac{(D-7)}{6 (4\pi)^{\frac{D}{2}}} \left\{\int_M\tau+2\int_{\pa M}\sum_{a=1}^{D-1}L_{aa}\right\}.
\end{split}
\end{equation}
\section{Casimir effect on the boundary of a $\boldsymbol{D}$-dimensional cavity}\label{sec5}

Now consider the Casimir effect on the shell $B$ which bounds  a cavity $M$. For simplicity, assume that $M$ is inside $\mathbb{R}^D$,   and it is star convex with respect to the the point $0$.
To   find  the electromagnetic Casimir free energy that gives rise to the Casimir stress on the shell $B$, we need to enclose the cavity $M$ in a much larger cavity $M_r$ of radius $r$. Specifically, we can let $$M_r=\{t\text{x}\in \mathbb{R}^D\,:\, 0\leq t\leq r, \text{x}\in M\}.$$ Let $A_r$ be the   annular region $\overline{M_r\setminus M}$. Then the boundary of $A_r$ is $B\cup B_r$, where $B_r$ is the boundary of $M_r$ given by
$$B_r=\{r\text{x}\in \mathbb{R}^D\,:\,  \text{x}\in B\}.$$
The  Casimir free energy of this configuration is given by
\begin{equation}\label{eq9_29_1}
E_{\text{Cas}}(B;b)=\lim_{r\rightarrow\infty}\left(E_{\text{Cas}}(M;b)+E_{\text{Cas}}(A_r;b)-E_{\text{Cas}}(M_r;b)\right),
\end{equation}i.e., the $r\rightarrow\infty$ limit of the sum of the Casimir free energies in $M$ and $A_r $ minus the Casimir free energy in $M_r$. For perfectly conducting conditions on $B$, $b=r$. For  infinitely permeable boundary conditions, $b=a$.

Using the result \eqref{eq4_25_2} of Section \ref{sec3},   we find that
\begin{equation}\label{eq4_5_3}\begin{split}
&E_{\text{Cas}}(M;b)+E_{\text{Cas}}(A_r;b)-E_{\text{Cas}}(M_r;b)\\=&\sum_{n=0}^{D-1}\frac{\Gamma(D+1-n)}{\Gamma\left(\frac{D-n}{2}\right)}\hat{c}_{n;b}\lambda^{n-D-1}-\frac{\psi(1) -\ln[\lambda\mu]}{2\sqrt{\pi}}\hat{c}_{D+1;b}+E_{\text{Cas}}^{\text{reg}}(M;b)+E_{\text{Cas}}^{\text{reg}}(A_r;b)-E_{\text{Cas}}^{\text{reg}}(M_r;b)+o(\lambda).\end{split}
\end{equation}The coefficients $\hat{c}_{n;b}$ for $n=0,1,\ldots,D-1$ and $D+1$ determine the divergence of the Casimir free energy. They are given by
\begin{equation*}
\hat{c}_{n;b}= c_{n;b}(M)+c_{n;b}(A_r)-c_{n;b}(M_r).
\end{equation*}
Notice that the coefficients $c_{n;b}(M)$ can be expressed as an integral over $M$ and an integral over the boundary of $M$. Since $M_r=A_r\cup M$, we find that for $\hat{c}_{n;b}$, the integrals over $M$ and $A_r$ cancel with the integral over $M_r$. On the other hand, since $\pa A_r=\pa M\cup\pa M_r $, we find that for $\hat{c}_{n;b}$, only the integral over the boundary of $M$ is left. Hence, it is always finite and independent of $r$.

From the result \eqref{eq4_25_3} of Section \ref{sec4}, we find that for absolute boundary conditions,
\begin{equation*}
\begin{split}
\hat{c}_{0;a}=&0,\\
\hat{c}_{1;a}=&\frac{(D-3)}{2(4\pi)^{\frac{D-1}{2}}}\text{vol}(B),\\
\hat{c}_{2;a}=&0.
\end{split}
\end{equation*}
For $\hat{c}_{2;a}$, the integrals over $\pa M$ cancel because $L_{aa}$ on the boundary $B$ as a boundary of $M$ and as a boundary of $A_r$ has opposite sign.

Similarly, for relative boundary conditions, \eqref{eq4_25_4} gives
\begin{equation*}
\begin{split}
\hat{c}_{0;r}=&0,\\
\hat{c}_{1;r}=&-\frac{(D-3)}{2(4\pi)^{\frac{D-1}{2}}}  \text{vol}(B),\\
\hat{c}_{2;r}=&0.
\end{split}
\end{equation*}

It is easy to see that $\hat{c}_{1;b}=0$ if and only if $D=3$. Therefore, if the dimension of the cavity $D$ is $\geq 4$, we find that the leading divergence is of order $\lambda^{-D}$. In this case, regularization is required.

When $D=3$, we have shown that  $\hat{c}_{0;b}=\hat{c}_{1;b}=\hat{c}_{2;b}=0$. In fact, it has been shown in \cite{7} that $\hat{c}_{4;b}=0$. From \eqref{eq4_5_3}, we find that the Casimir free energy is   finite when $\lambda\rightarrow 0^+$ if and only if $\hat{c}_{0;b}=\ldots=\hat{c}_{D-1;b}=\hat{c}_{D+1;b}=0$. Therefore we find that in $D=3$ dimensions, no regularization is required for the Casimir free energy. This is the main result obtained in \cite{7}. In this work, we find that the Casimir free energy always require regularization when $D>3$. This probably explain why physics in $(3+1)$-dimensions  are special.

Let us look at the regularized Casimir free energy. One can argue that the limit
\begin{equation*}
\lim_{r\rightarrow\infty}\Bigl(\zeta_{T;b}'(0;M)+\zeta_{T;b}'(0;A_r)-\zeta_{T;b}'(0;M_r)\Bigr)
\end{equation*}is finite. Following from  \eqref{eq9_29_1} and \eqref{eq10_17_8}, the regularized Casimir free energy of the shell $B$, denoted by $E_{\text{Cas}}^{\text{reg}}(B;b)$, is defined as
\begin{equation*}\begin{split}
E_{\text{Cas}}^{\text{reg}}(B;b)=&-\frac{T}{2}\lim_{r\rightarrow\infty}\Bigl\{\zeta_{T;b}'(0;M)+\zeta_{T;b}'(0;A_r)-\zeta_{T;b}'(0;M_r)+[\ln\tilde{\mu}^2]
\bigl(\zeta_{T;b}(0;M)+\zeta_{T;b}(0;A_r)-\zeta_{T;b}(0;M_r)\bigr)\Bigr\}\\
=&-\frac{T}{2}\lim_{r\rightarrow\infty}\Bigl\{\zeta_{T;b}'(0;M)+\zeta_{T;b}'(0;A_r)-\zeta_{T;b}'(0;M_r) \Bigr\}-\frac{\ln\tilde{\mu}^2}{4\sqrt{\pi}}\hat{c}_{D+1,b}.
\end{split}\end{equation*}This is free of ambiguities if and only if $\hat{c}_{D+1;b}$ is zero, which is known to be the case when $D$ is odd \cite{5}.

In the high temperature limit, the asymptotic expansion of the regularized Casimir free energy is given by
\begin{equation}\label{eq10_18_3}
\begin{split}
  E_{\text{Cas}}^{\text{reg}}(B;b)\sim &-\frac{1}{\sqrt{\pi}}\sum_{n=0}^{D-1}2^{D-n} \Gamma\left(\frac{D-n+1}{2}\right)\zeta_R(D-n+1) \hat{c}_{n;b}T^{D-n+1}-T\left(\hat{c}_{D;b}\ln T+\frac{Q_b}{2} \right)\\
&+\frac{\Bigl(1+\psi(1)+\ln (2\pi )+\ln( T/\mu)\Bigr)}{2\sqrt{\pi}}\hat{c}_{D+1;b}- \sum_{n=D+2}^{\infty}\frac{1}{(2\pi)^{n-D}} \Gamma\left(\frac{n-D}{2}\right)\zeta_R(n-D) \hat{c}_{n;b}\frac{1}{T^{n-D-1}},
\end{split}
\end{equation}
where
\begin{equation*}
Q_b=\lim_{r\rightarrow \infty}\Bigl(\zeta_{b}'(0;M)+\zeta_b'(0;A_r)-\zeta_b'(0;M_r)\Bigr).
\end{equation*}
As is discussed in \cite{8,9}, the Casimir free energy has to be renormalized to remove terms of order $T^2,\ldots, T^{D+1}$  in the high temperature limit. Therefore, the renormalized (physical) Casimir free energy is given by
\begin{equation*}\begin{split}
E_{\text{Cas}}^{\text{ren}}(B;b)=&-\frac{T}{2}\lim_{r\rightarrow\infty}\Bigl\{\zeta_{T;b}'(0;M)+\zeta_{T;b}'(0;A_r)-\zeta_{T;b}'(0;M_r)+[\ln\tilde{\mu}^2]
\bigl(\zeta_{T;b}(0;M)+\zeta_{T;b}(0;A_r)-\zeta_{T;b}(0;M_r)\bigr)\Bigr\}\\&+\frac{1}{\sqrt{\pi}}\sum_{n=0}^{D-1}2^{D-n} \Gamma\left(\frac{D-n+1}{2}\right)\zeta_R(D-n+1) \hat{c}_{n;b}T^{D-n+1},
\end{split}\end{equation*}which involves the coeffcients $\hat{c}_{n;b}$ for $0\leq n\leq D-1$. When $D=3$, the last term is zero since $\hat{c}_{0;b}=\hat{c}_{1;b}=\hat{c}_{2;b}=0$ and therefore no renormalization is needed.

The leading term of the physical Casimir free energy is
\begin{equation*}
-T\left(\hat{c}_{D;b}\ln T+\frac{Q_b}{2} \right).
\end{equation*}It has a term $T\ln T$ with coefficient $-\hat{c}_{D;b}$.

In summary, we find that the coefficients $\hat{c}_{n;b}$ for $0\leq n\leq D-1$ is related to the divergence of the zero temperature Casimir energy, and also appear in the renormalization of the Casimir free energy. The coefficient $\hat{c}_{D;b}$ gives rise to a term proportional to $T\ln T$ in the high temperature limit. The vanishing of the coefficient $\hat{c}_{D+1;b}$ is required for the regularized Casimir free energy to be well-defined.

\section{Conclusion}
In this work, we consider the electromagnetic Casimir effect acting on the boundary of a $D$-dimensional cavity. We show that the perfectly conducting and infinitely permeable boundary conditions correspond respectively to relative and absolute boundary conditions for one-forms. Using exponential cut-off method, we investigate the divergence structure of the Casimir free energy, and show that they are related to heat kernel coefficients of Laplace operators on one-forms. After some cancelations between the divergences inside and outside the cavity, we find that the leading term of the divergence of the Casimir free energy is equal to  a constant times $(D-3)$ times the volume of the boundary of the cavity. This shows that when the dimension $D$ is larger than three, the divergences do not cancel out and regularization is always required. When $D=3$, it has been proved in \cite{7} that all the divergences always cancel out. 

We also investigate the high temperature asymptotic behavior of the Casimir free energy. It is shown that the coefficients of the terms of order $T^2, T^3,\ldots, T^{D+1}$ are multiples of the first $D$ heat kernel coefficients. As the case of the divergences, these terms all vanish if and only if $D=3$. When $D>3$, renormalization is required to remove these terms.  

\begin{acknowledgments}\noindent
  This work is supported by the Ministry of Higher Education of Malaysia  under   FRGS grant FRGS/1/2013/ST02/UNIM/02/2. I would like to thank K. Kirsten for the helpful discussions.
\end{acknowledgments}

\end{document}